# Targeting the Synergistic Interaction of Pathologies in Alzheimer's Disease: Rationale and Prospects for Combination Therapy

Author: She Xutong

Affiliation: Nanjing University of Science and Technology

Corresponding Author: She Xutong

Email: shexutong@njust.edu.cn

ORCID iD: Xutong She (0009-0003-3242-0202) - ORCID

## Abstract

Alzheimer's disease (AD) persists as a paramount challenge in neurological research, characterized by the pathological hallmarks of amyloid-β (Aβ) plaques and neurofibrillary tangles composed of hyperphosphorylated tau. This review synthesizes the evolving understanding of AD pathogenesis, moving beyond the linear amyloid cascade hypothesis to conceptualize the disease as a cross-talk of intricately interacting pathologies, encompassing Aβ, tau, and neuroinflammation. This evolving pathophysiological understanding parallels a transformation in diagnostic paradigms, where biomarker-based strategies—such as the AT(N) framework—enable early disease detection during preclinical or prodromal stages. Within this new landscape, while anti-Aβ monoclonal antibodies (e.g., lecanemab, donanemab) represent a breakthrough as the first disease-modifying therapies, their modest efficacy underscores the limitation of single-target approaches. Therefore, the author explore the compelling rationale for combination therapies that simultaneously target Aβ pathology, aberrant tau, and neuroinflammation. Looking forward, I emphasize emerging technological platforms—such as gene editing and biophysical neuromodulation—in advancing precision medicine. Ultimately, the integration of early biomarker detection, multi-target therapeutic strategies, and AI-driven patient stratification charts a promising roadmap toward fundamentally altering the trajectory of AD. The future of AD management will be defined by preemptive, biomarker-guided, and personalized combination interventions.

Keywords: Alzheimer's disease, amyloid-β, tau pathology, neuroinflammation, combination therapy, multi-target therapy, precision medicine, biomarkers

# 1. Definition

Aging is a normative physiological process characterized by a progressive, systemic decline in tissue and organ functions over time. In the cognitive domain, typical aging is associated with subtle, non-progressive alterations that do not compromise autonomy in activities of daily living.[1] In contrast, dementia is not an inherent consequence of normal aging,[2] but an qualitatively distinct, acquired clinical illness. It is defined by impairment in at least two cognitive domains,[3] such as language, executive functioning, memory, attention, or visuospatial abilities,[4,5] with deficits severe enough to significantly interfere with daily living activities, social interactions, or interpersonal relationships.[6] Illustrative manifestations of dementia include failure to recognize familiar individuals, disorientation in previously familiar environments,[5] abrupt mood lability without clear precipitant,[7] and marked personality changes.[8]

Alzheimer's disease (AD), the most prevalent cause of dementia, is a progressive and irreversible neurodegenerative disorder.[9,10] Its hallmark neuropathological features include extracellular deposition of β-amyloid (Aβ) plaques,[11] intraneuronal accumulation of hyperphosphorylated tau as neurofibrillary tangles,[12] and consequent neuronal loss,[13] synaptic degeneration, and cerebral atrophy,[14] with predilection for the hippocampus and medial temporal lobe structures.[15-17]

Clinically, AD typically evolves through three stages—mild, moderate, and severe.[6,10,18] The mild stage (approximately 0–3 years) is characterized primarily by episodic memory impairment, manifesting as frequent forgetting of recent events,[19] repetitive questioning, and decline in instrumental activities of daily living.[10,20] The moderate stage is marked by escalating behavioral disturbances (e.g., inability to complete complex tasks such as cooking),[21] language impairments (e.g., anomia and word-finding difficulty), and visuospatial deficits (e.g., getting lost in familiar settings).[22] In the severe stage, individuals lose independence, develop profound behavioral abnormalities (e.g., agitation, apathy), and exhibit global cognitive failure.[23] The overall disease course typically spans 5–10 years,[24] with mortality most often attributable to complications such as infection or malnutrition,[25] which include hypercholesterolemia, hypertension, diabetes, obesity, depression, and cardiovascular diseases;[26] and complications ensuing from disease progression—such as thrombosis, immobility, dysphagia, malnutrition, and pneumonia (pulmonary infection)—both of which significantly contribute to an increased mortality risk.

Risk factors for dementia **Figure 1** include advancing age, a positive family history, prior head injury, and lower educational attainment.[27,28] The genetic architecture of dementia encompasses rare, early-onset (before age 60) familial forms associated with autosomal-dominant mutations on chromosomes 1, 14, and 21.[29-31] Genes implicated in autosomal-dominant early-onset AD include amyloid precursor protein (APP), Presenilin-1 (PSEN1), and Presenilin-2 (PSEN2).[32] In late-onset Alzheimer's disease (AD), the apolipoprotein E (APOE) locus on chromosome 19 is a major susceptibility factor: the ε4 allele confers a dose-dependent increase in risk and lowers the age at onset, whereas the ε2 allele may be protective.[33,34]

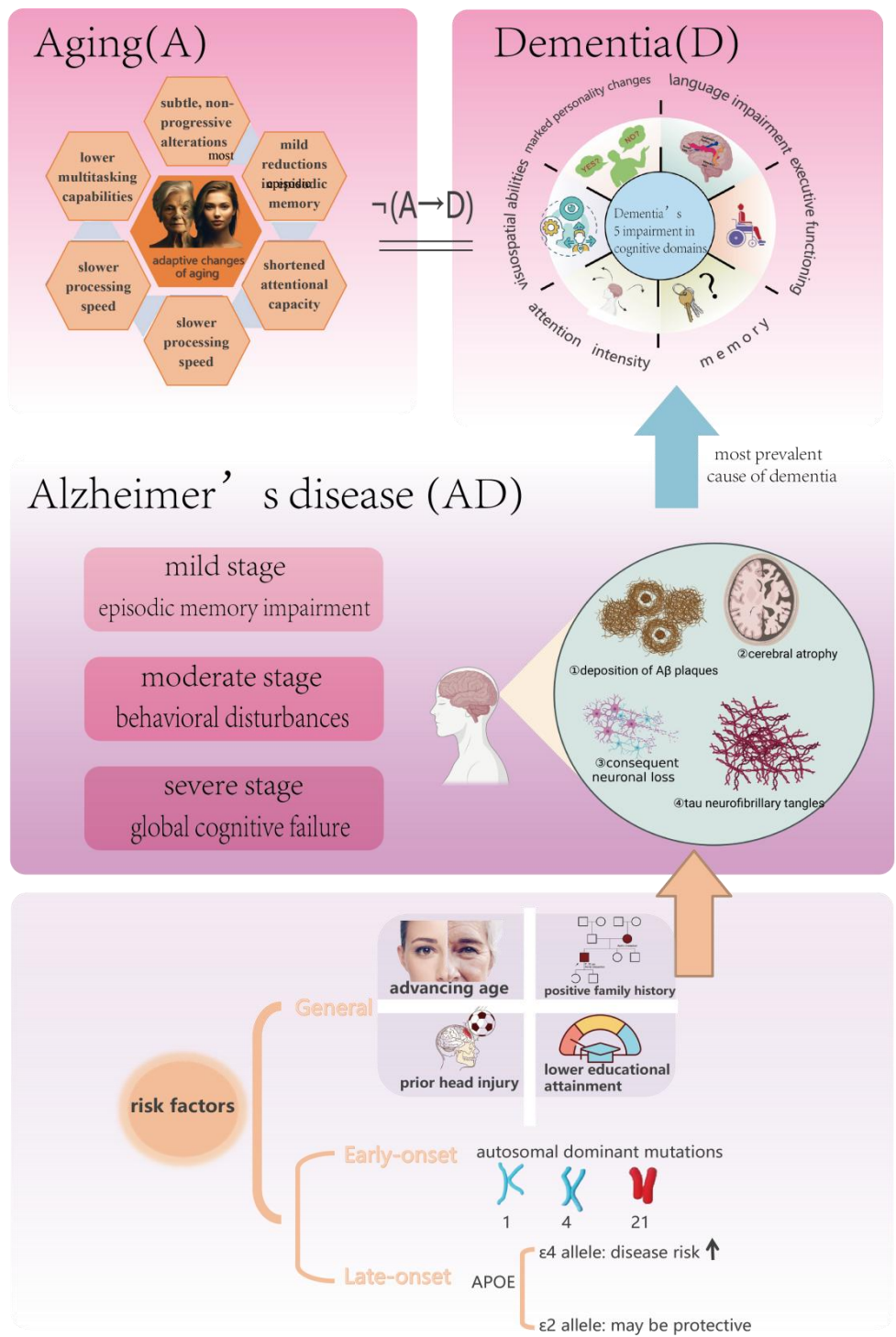

**Figure 1** Schematic Illustration of Aging,[1] Dementia,[3] and Alzheimer's Disease:[6] From Cognitive Changes to Pathological Mechanisms and Risk Factors

Diagnosis of AD integrates multiple modalities. Core elements include standardized neuropsychological assessment—The Montreal Cognitive Assessment (MoCA),[35] to document impairments across relevant cognitive domains and structured interviews with informants (family members or caregivers) to ascertain functional decline and symptom trajectory.[10,36,37] Concurrently, a comprehensive evaluation is required to exclude alternative or contributory etiologies, including thyroid dysfunction, neurosyphilis, metabolic and systemic disturbances (e.g., renal insufficiency, electrolyte derangements, diabetes), heavy metal toxicity (e.g., lead, mercury), and anemia, among others.[10,37] Neuroimaging—typically CT or MRI, and when indicated PET—are commonly used to detect structural and functional brain activity in vivo, supporting the exclusion of other intracranial pathologies (e.g., cerebrovascular disease, traumatic sequelae, neoplasms) and may provide supportive evidence for AD.[38]

## 2. Social Cost

Dementia is an acquired syndrome characterized by decline in memory and other cognitive domains sufficient to impair daily functioning in an otherwise alert individual, and it remains a pressing global health concern.[2-4,6] The World Health Organization's 2022 blueprint for dementia research estimates that approximately 55.2 million people worldwide are affected. Earlier estimates indicated that roughly 40 million individuals were living with Alzheimer's disease (AD) in 2016, with projections suggesting a doubling of cases approximately every two decades.[39]

Although the proportion of older adults is lower in many developing countries than in Western and Northern Europe, reported incidence and prevalence of AD can be higher in these settings.[40] Epidemiological data from high-income countries indicate a prevalence of approximately 4%–8% among individuals aged 65 and older.[41] In China, studies led by Professor Jia Jianping report a prevalence of approximately 3%–7%, with higher rates in women than in men;[42] current estimates suggest 6–8 million people in China are living with AD. The World Alzheimer Report 2013 attributes 50%–75% of dementia cases to AD. Prevalence increases steeply with age: it roughly doubles every 6.1 years, reaching 20%–30% by age 85.[43] Globally, the World Alzheimer Report 2018 (Alzheimer's Disease International) estimates at least 50 million people living with dementia, with projections of 152 million by 2050, of whom approximately 60%–70% are expected to have AD.[44]

Significant socioeconomic and caring costs are imposed by AD. According to the World Alzheimer Report 2015, global dementia-related costs were estimated at US$606.7 billion in 2010 and US$817.9 billion in 2015 (adjusted for prevalence and World Bank country classifications).[45] Advancing research and improving treatment and care pathways for AD are therefore critical to mitigating societal and familial burdens.

## 3. Mechanisms

Numerous hypotheses have been proposed to explain the pathogenesis of Alzheimer's disease

(AD), yet no unified theory has emerged, reflecting the disorder's substantial biological and clinical complexity. AD is broadly classified into two forms: familial and sporadic.[46] Familial Alzheimer's disease (FAD), comprising approximately 1–5% of cases, arises from pathogenic variants in autosomal-dominant genes—APP, PSEN1, and PSEN2—and typically presents as early-onset AD (EOAD).[47] FAD exhibits clear familial aggregation, often affecting multiple first-degree relatives.[48] Although FAD represents less than 1% of all AD overall,[32] it is of outsized importance for elucidating the underlying pathogenesis of AD, particularly the amyloid pathway.[11]

In contrast, sporadic Alzheimer's disease (SAD), accounting for over 95% of cases, does not follow a monogenic autosomal-dominant inheritance pattern.[11] Its pathogenesis reflects the interplay of polygenic susceptibility—most notably the APOE ε4 allele[33]—with modifiable risk factors, including cardiometabolic conditions, physical inactivity, and limited cognitive reserve; symptom onset most commonly occurs after age 65.[49,50] Genome-wide association studies and meta-analyses have identified numerous risk loci for SAD,[51] implicating pathways related to immune function, lipid metabolism, amyloid-β processing and deposition,[52] tau pathology (neurofibrillary tangle formation),[53] and endocytosis,[54] while leaving a substantial proportion of heritability unexplained.[55]

Non-genetic, modifiable determinants—spanning lifestyle behaviors, psychosocial variables, environmental exposures, and AD-associated comorbidities—also contribute meaningfully to risk, likely by modulating biological pathways and interacting with genetic susceptibility.[55-58] This multifactorial architecture complicates attribution of a singular causal pathway for clinical AD.[59-61] Further, AD encompasses typical and atypical clinical subtypes with diverse phenotypic presentations, and its neuropathology is similarly multifaceted, featuring Aβ plaques, neurofibrillary tangles, synaptic and neuronal loss, and neuroinflammation.[62,63]

In sum, the pronounced heterogeneity of AD arises from multidimensional variation across etiologic drivers, clinical phenotypes, and neuropathologic profiles.**Figure 2** As a result, constructing an integrative framework that seamlessly links genetic underpinnings, molecular mechanisms, and clinical expression remains formidable. Current methodological and evidentiary limitations further constrain a comprehensive understanding of AD pathophysiology.[39]

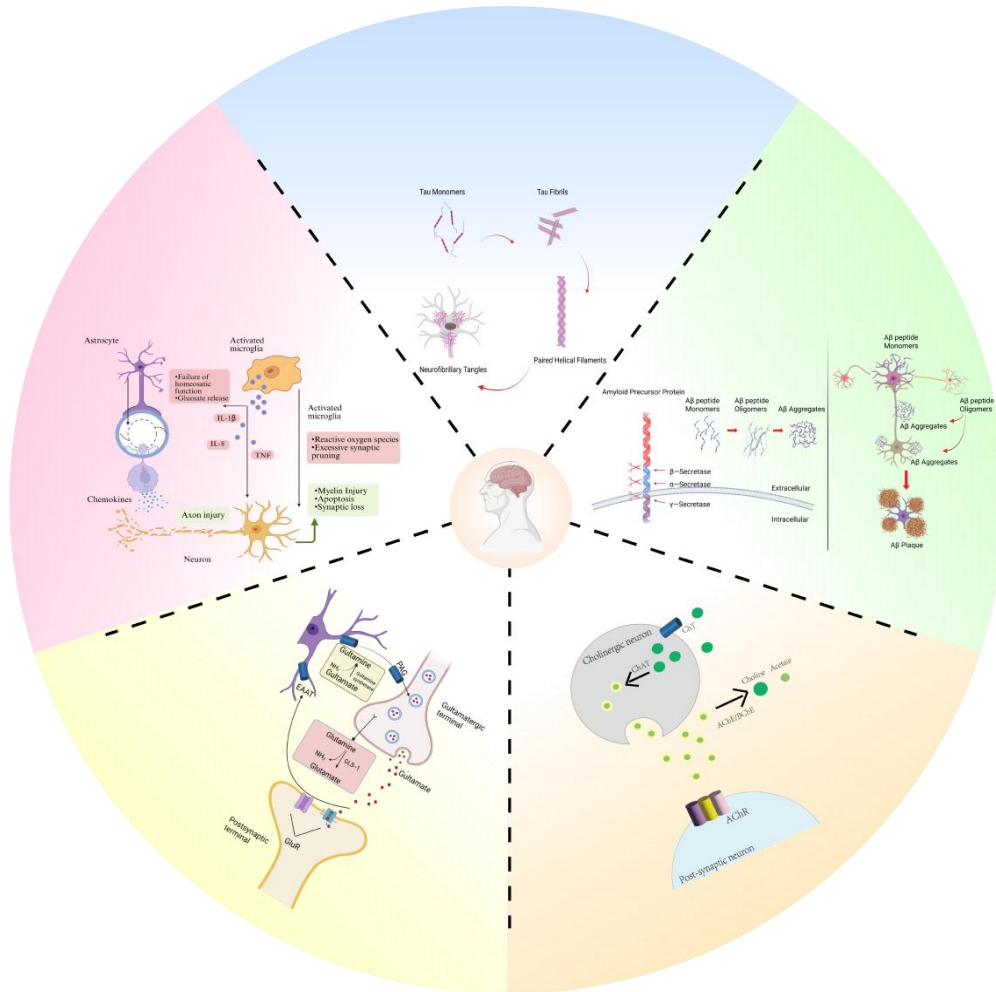

**Figure 2** Diagram for the pathogenesis of AD, including the the amyloid hypothesis,[64] the inflammatory hypothesis,[65] cholinergic hypothesis,[66] the tau protein hypothesis,[67] the glutamatergic hypothesis[68]

### 3.1.1 Amyloid hypothesis

The accumulation of amyloid-β (Aβ) peptides is one of the earliest detectable pathological events in Alzheimer's disease (AD), preceding clinical symptoms by years or even decades.[69] This temporal primacy established the amyloid cascade hypothesis, which posits Aβ as a key driver of the downstream neurodegenerative cascade. APP is processed via the amyloidogenic pathway through sequential cleavage by β-site APP–cleaving enzyme 1 (BACE1) and the γ-secretase complex, producing Aβ peptides (**Figure 3**). Aβ peptides, particularly the Aβ42 isoform, are derived from the sequential cleavage of amyloid precursor protein (APP) by β- and γ-secretases. The hydrophobic Aβ42 is highly aggregation-prone, initially forming soluble neurotoxic oligomers that impair synaptic function, and ultimately depositing into insoluble fibrils and plaques.[70]

The intramembrane proteolysis mediated by γ-secretase is inherently distributive and yields a heterogeneous mixture of Aβ isoforms, most prominently Aβ40 and Aβ42.[71,72] Aβ40, a 40–amino acid peptide, is relatively soluble and less prone to aggregation, whereas Aβ42, extended by two hydrophobic residues at its C-terminus (isoleucine and alanine), exhibits markedly increased hydrophobicity and aggregation propensity.[11,73] Aβ42 readily self-associates, initially forming small soluble oligomers that nucleate the assembly of insoluble fibrils, which are potent disruptors of synaptic function.[74] These fibrils accumulate into mature amyloid fibers and deposit extracellularly as amyloid plaques (e.g., neuritic/senile plaques), often accompanied by neuroinflammatory changes.

Mounting evidence implicates soluble Aβ oligomers—derived from the aggregation of Aβ monomers—as the principal neurotoxic species in Alzheimer's disease pathogenesis.[64] These oligomers from the aggregation of Aβ monomers can propagate throughout the brain parenchyma, and directly bin to neuronal surface receptors including NMDA and insulin receptors. This binding triggers a cascade of neurotoxic events: disruption of membrane integrity,[75] aberrant innate immune activation, perturbation calcium homeostasis,[76] mitochondrial impairment,[77] oxidative stress,[78] and ultimately, synaptic dysfunction.[79] As the aggregation process continues, these soluble oligomers assemble into larger, structured protofibrils, which ultimately mature into insoluble fibrils that deposit as amyloid plaques. In contrast to the direct toxicity of oligomers, plaques exhibit relatively limited direct neurotoxicity.[80] Their primary contribution to pathogenesis lies in functioning as a reservoir for soluble oligomers and, crucially, in perpetuating a chronic and destructive neuroinflammatory response through the sustained activation of surrounding microglia and astrocytes.[81] This is a self-reinforcing cycle that ultimately exacerbates Aβ pathology through mechanisms including the release of Apoptosis-Associated Speck-like Protein Containing a CARD (ASC) specks.[82]

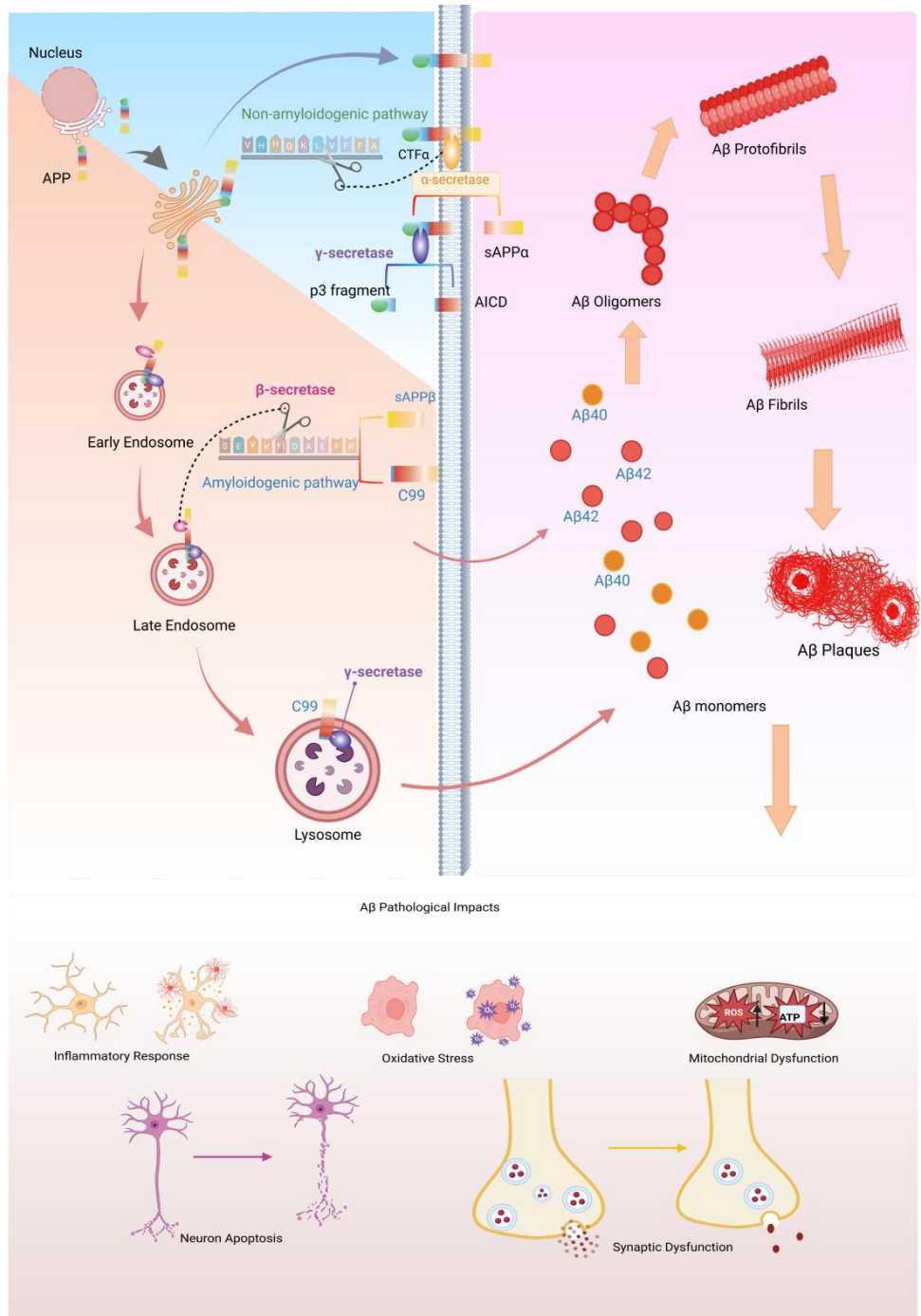

**Figure 3**  APP Cleavage Pathways in Alzheimer's Disease: Amyloidogenic[83] vs. Non-Amyloidogenic Routes and Their Sequelae of Aβ Aggregation and Pathological Impacts

Evidence indicates that Alzheimer's disease (AD) reflects a dynamic imbalance between the production and clearance of amyloid-β (Aβ),[84,85] particularly Aβ42—arising from either excessive generation or impaired removal.[86] In familial AD (FAD), an inherited form of Alzheimer's disease caused by mutations in genes such as APP, PSEN1, and PSEN2, enhanced aggregation of Aβ42 promotes the formation of neurotoxic oligomers that ultimately deposit as amyloid plaques and exert multifaceted neurotoxicity, including synaptic dysfunction, disruption of neuronal membranes, oxidative stress, mitochondrial impairment, chronic neuroinflammation, facilitation of tau hyperphosphorylation, and cerebrovascular injury.[87] Notably, while mutations in the PSEN1 gene can contribute to Aβ accumulation through multiple mechanisms in familial AD (FAD),[52,88-90] it is important to highlight that FAD-associated mutations do not invariably increase absolute Aβ42 levels or the Aβ42/Aβ40 ratio.[70,91]

In sporadic AD (SAD), a non-inherited form with elusive etiology, plaque formation is more complex and is thought to be driven predominantly by inefficient Aβ clearance. A pivotal molecule in this context is apolipoprotein E (ApoE), a lipid-transport protein primarily produced in the liver and the brain. It plays a critical role in regulating the clearance and aggregation of amyloid-β (Aβ). Under physiological conditions, ApoE delivers lipids to neurons, thereby supporting synaptic integrity, facilitating synaptic formation and repair. In the context of Alzheimer's disease (AD) pathology, however, the specific ApoE ε4 isoform becomes dysfunctional and exhibits an impaired ability to clear Aβ, primarily due to its lower binding affinity for Aβ and its significantly less efficient mediation of Aβ clearance through the blood-brain barrier and cellular phagocytic pathways compared to other isoforms, like ApoE ε 2 and ApoE ε 3. Among those three common alleles, APOE ε4 allele is the strongest common genetic risk factor for SAD- carriage of one ε4 allele increases AD risk approximately two to three folds, and two ε4 alleles confer an estimated 12-fold increase.[92] APOE protein is detectable within neuritic plaques, and individuals with ε4 alleles exhibit greater cerebral Aβ plaque burden, supporting the hypothesis that APOE ε4 adversely affects Aβ clearance mechanisms.[93]

Clinically, the development of anti-Aβ monoclonal antibodies—such as aducanumab,[94] lecanemab,[95] and donanemab[96] represents a breakthrough, demonstrating robust plaque clearance and modest slowing of cognitive decline in early AD. However, their benefits are partial, and their efficacy wanes in later disease stages.[97,98] The more fundamental challenge, however, is that various late-stage trials of Aβ-lowering therapies have failed to demonstrate meaningful cognitive benefit, particularly in moderate-to-severe disease.[99,100] These repeated failures have revealed A core lesson: therapeutic intervention must be significantly advanced to the early stage of the disease cascade, and Aβ clearance as a single strategy is far from sufficient. This conclusion is bolstered by clinicopathological observations showing an imperfect correspondence between Aβ plaque burden and cognitive impairment.[101] Some older adults harbor abundant Aβ plaques at autopsy despite preserved premortem cognitive function,[102] whereas others with demonstrable cognitive impairment show relatively modest plaque burden.[103,104]. This discrepancy has given rise to alternative hypotheses which reframe Aβ aggregation as a downstream consequences or secondary phenomena within a broader pathological cascade,[50,105] arguing that it is necessary but not sufficient to drive the disease.

According to this view, the onset of symptomatic AD obligatorily requires convergence with additional factors—such as tau pathology, neuroinflammation, vascular injury, and variability in cognitive reserve.[106] This multifactorial nature of AD provides a compelling rationale for pursuing combination therapies from the earliest stages.

Aβ pathology initiates the disease cascade and creates a permissive environment for the spread of tau and the ignition of neuroinflammation. Conquering AD will not come from defeating Aβ alone, but also from leveraging early Aβ detection to identify at-risk individuals and deploying combination therapies that concurrently mitigate Aβ, tau, and inflammatory pathologies. Given the central role of tau pathology in disease progression, elucidating the temporal ordering[107] and mechanistic interplay between tau and Aβ[108] in AD warrants rigorous investigation.[109-111]

### 3.1.2. Tau protein hypothesis

The tau hypothesis posits that aberrant hyperphosphorylation, mislocalization, and aggregation of the microtubule-associated protein tau into neurofibrillary tangles (NFTs) precipitate collapse of the neuronal microtubule network, impair axonal transport, disrupt synaptic function, and ultimately drive neuronal death[112,113]—constituting principal mechanisms underlying neurodegeneration and cognitive decline in Alzheimer's disease (AD).[113] Tau is predominantly expressed in neuronal axons,[114-116] where it binds to and stabilizes microtubules,[117] key elements of the cytoskeleton that function as intracellular "tracks" for bidirectional transport of nutrients, organelles (e.g., mitochondria and vesicles), and signaling molecules between the soma and axon terminals.[118,119] Unlike Aβ, which accumulates early, the spread of tau pathology from the medial temporal lobe to the neocortex closely tracks the progression of clinical symptoms, making it a powerful correlate—and likely executor—of cognitive decline.

Physiologically, tau stabilizes microtubules in neuronal axons. Pathologically, it becomes hyperphosphorylated, detaches from microtubules,[112,120,121] assemble into insoluble fibril,[67,122-124] and aggregates into insoluble NFTs—a hallmark lesion of the AD brain.[113,125] These processes culminate in neuronal injury and cell death. Critically, pathological tau propagates trans-synaptically through brain networks and instigates neuroinflammation by activating microglia and astrocytes, creating a self-reinforcing cycle of damage.

The strong spatiotemporal correlation between tau deposition and clinical symptoms positions tau pathology as a crucial biomarker for disease staging and early diagnosis.

Multiple lines of evidence, including clinical and experimental ones, support the tau hypothesis:

**Clinicopathological correlation:** Braak staging, a system used to assess the progression of neurodegenerative diseases, which maps the stereotyped spatiotemporal progression of tau pathology, correlates strongly with the severity of cognitive impairment—more robustly than regional amyloid-β plaque distribution.[14,126]

**In vivo imaging:** Tau-PET tracers (e.g., flortaucipir, MK-6240, PI-2620) visualize tau deposition in living patients; signal intensity and topography show high concordance with cognitive deficits and with regional patterns of cortical atrophy, exceeding the correspondence observed with Aβ-PET.[127,128]

**Experimental models:** Transgenic mice expressing mutant human tau (e.g., P301L) develop hallmark features—tau aggregation, neuronal loss, brain atrophy, and cognitive impairment—recapitulating core aspects of AD-related neurodegeneration.[129,130]

Despite its clear role in symptom manifestation, targeting tau alone has proven challenging. Therapeutic strategies—including kinase inhibitors, aggregation inhibitors, and immunotherapies—have thus far demonstrated limited clinical efficacy. This underscores that tau pathology, while closely linked to symptoms, does not operate in isolation. Its progression is modulated by interactions with Aβ, neuroinflammation, and genetic factors such as APOE ε4 and TREM2 variants. For instance, emerging evidence indicates that women with the APOE ε4/ε4 genotype exhibit heightened susceptibility to Aβ-driven tau accumulation,[131,132] highlighting the complex interplay between core pathologies and individual risk factors. Additionally, mediated by impaired microglial function, carriers of rare TREM2[133] risk variants and/or APOE ε4 homozygosity[134] display more extensive dissemination of tau pathology[135] from the EC to neocortical regions, though molecular crosstalk between microglial activation and tau pathogenicity is not yet defined.[136] These observations underscore a critical interplay among genetic determinants, microglial function, and sex-specific factors in AD progression, supporting precision-medicine strategies tailored to individual risk profiles.

Beyond genetic and inflammatory factors, metabolic disturbances also contribute to tau pathology.[137] Converging research highlights that mitochondrial dysfunction, influenced by mitochondrial DNA (mtDNA) variation, can promote aberrant tau phosphorylation via stress-activated kinase pathways (e.g., GSK3B).[138] These findings implicate mtDNA in metabolic-sensing pathways that enhance tau pathology, providing a novel mechanistic framework for diagnostic and therapeutic approaches targeting mitochondrial integrity.[138]

In summary, tau pathology serves as a key nexus linking early Aβ deposition to later-stage cognitive decline and neurodegeneration. Its strong correlation with symptoms makes it an invaluable biomarker for early diagnosis and staging. However, the limited success of tau-specific therapeutics reinforces the necessity of combination strategies. Effective AD treatment will likely require concurrently mitigating Aβ seeding, halting tau propagation, and dampening the neuroinflammatory response, tailored to an individual's specific pathological and genetic profile.

### 3.1.3 Interplay between Aβ and Tau

The Aβ cascade hypothesis states that Aβ, deposited in the form of neuroinflammatory plaques, induces AD by damaging neuronal cells.[139] Aβ facilitates the development of AD and initiates a deleterious cascade involving tau pathology and neurodegeneration.[140] Nonetheless, extensive studies focused solely on the neurotoxicity of Aβ or tau have not shown significant efficacy in the treatment of AD.[141,142] Therefore, focusing solely on the role of Aβ or tau in AD lesions while ignoring the interaction between Aβ and tau may not be entirely correct. While historically studied as parallel pathways, emerging evidence underscores their synergistic cooperation, wherein amyloid accumulation may facilitate the broader dissemination of tau,[143,144] and toxic state of tau can enhance Aβ toxicity via a feedback loop.[145,146]

This deleterious feedback loop is critically mediated by the action of Aβ on tau. Specifically, Aβ orchestrates tau hyperphosphorylation primarily by activating key cellular kinases.[110] Soluble Aβ oligomers engage neuronal surface receptors such as NMDA and insulin receptors, triggering intracellular signaling cascades that lead to calcium influx and oxidative stress. These events activate stress-responsive kinases, including glycogen synthase kinase-3β (GSK-3β) and cyclin-dependent kinase 5 (CDK5).[147] GSK-3β, in particular, is a major tau kinase whose activity is upregulated by Aβ-induced disruption of calcium homeostasis and mitochondrial function.[148] Once activated, GSK-3β phosphorylates tau at multiple epitopes (e.g., Ser396,Thr231), reducing its affinity for microtubules and promoting its dissociation.[149] This facilitates tau misfolding,[150] oligomerization,[151] and eventual aggregation into neurofibrillary tangles (NFTs).[152,153] Therefore, Aβ-driven neuroinflammation—characterized by microglial release of pro-inflammatory cytokines such as TNF-α and IL-1β—further enhances kinase activation,[154] creating a feed-forward loop that accelerates tau pathology,[155] and reduce in Aβ clearance and accumulation in the brain.[156]

Conversely, tau pathology actively exacerbates Aβ accumulation and toxicity. Mislocalized and hyperphosphorylated tau impairs axonal transport,[157] including the trafficking of amyloid precursor protein (APP) and its processing enzymes.[158] This disruption favors amyloidogenic APP cleavage by BACE1 and γ-secretase, increasing Aβ production.[159] In addition, tau aggregates contribute to synaptic failure and neuronal hyperexcitability, which elevate neuronal activity and promote Aβ release. Furthermore, extracellular tau species—released from degenerating neurons—can act as "seeds" that promote Aβ aggregation by cross-seeding or by stabilizing toxic Aβ oligomers.[160]

This reciprocal interaction between Aβ and tau establishes a destructive cycle that drives AD progression. The initial Aβ deposition creates a permissive environment for tau phosphorylation and spread,[161,162] while tau pathology in turn promotes Aβ release,[160] compromises clearance, and heightens neuronal vulnerability. This interplay underscores the limitation of monotherapeutic approaches and highlights the necessity of combination strategies that concurrently target both pathologies. Future interventions must account for this dynamic crosstalk,[108] leveraging biomarkers to identify patients with co-existing Aβ and tau pathologies for tailored, multi-target therapy.

## 3.2 Neuroinflammation: An Amplifier of Pathology

### 3.2.1 Neuroinflammation

Neuroinflammation, characterized by chronic activation of microglia and astrocytes alongside pro-inflammatory mediator release, represents a sustained immune response within the Central Nervous System(CNS).[163,164] In Alzheimer's disease (AD), this response is triggered by amyloid-β (Aβ) plaques, tau tangles, neuronal injury, or ancillary insults (e.g., infection, toxins, metabolic disturbances)—thereby shifting from a protective mechanism to a pathogenic process characterized by sustained release of neurotoxic mediators, synaptic and neuronal injury, and acceleration of disease progression.[165,166] Critically, neuroinflammation is not merely a consequence of pathology but an active contributor that interconnects the amyloid and tau cascades.

Microglia, the brain's resident immune cells, play a dual role. Under physiological conditions, they support synaptic homeostasis.[167] In AD, they are activated by Aβ[168] and attempt to clear plaques via receptors like TREM2 (Triggering Receptor Expressed on Myeloid cells 2). TREM2 is a key receptor on the surface of microglia. In the AD brain, it senses "danger signals" such as Aβ plaques and neuronal debris and binds to ligands like ApoE and Aβ. However, chronic activation leads to phagocytic dysfunction, sustained release of pro-inflammatory cytokines (e.g., IL-1β, TNF-α),[168] and NLRP3 inflammasome activation,[169] which paradoxically promotes further Aβ aggregation.[82] Furthermore, these inflammatory mediators directly exacerbate tau pathology; for instance, TNF-α can activate GSK-3β,[155] a kinase that drives tau hyperphosphorylation.[82] [170] This creates a self-reinforcing cycle where Aβ and tau jointly sustain a neurotoxic environment, and inflammation, in turn, facilitates tau propagation and synaptic damage. Excessive activation of the complement cascade may also drive aberrant microglia-mediated synaptic pruning, further aggravating neurodegeneration.[163]

Microglial activation progresses through distinct states characterized by discrete transcriptional signatures—early phases enriched for cell-proliferation programs and later phases for immune-response pathways.[171] TREM2 functions as a critical regulator: under physiological or early pathological conditions, it supports tau clearance;[172] however, at advanced stages, altered TREM2 signaling may paradoxically facilitate tau dissemination, complicating therapeutic approaches. Consequently, indiscriminate "anti-inflammatory" strategies risk blunting beneficial microglial functions. A more nuanced objective is immunomodulation—reprogramming microglia to suppress deleterious outputs (e.g., excessive pro-inflammatory cytokine release) while preserving or enhancing protective activities (e.g., phagocytosis and debris clearance).[173]

Given the complex role of TREM2 protein and its potential as a promising target for the prevention of Alzheimer's disease (AD), it is essential to investigate how TREM2 mediates the activation, migration, and clearance of Aβ.

Upon ligand of TREM2 binding, the ITAM motif of its partner DNAX-activation protein 12 (DAP12) becomes phosphorylated, leading to the recruitment and activation of the cytoplasmic kinase Syk. This, in turn, activates the PI3K-Akt pathway, the PLCγ pathway, and the Vav-Rac pathway. The PI3K-Akt pathway promotes cell survival and proliferation. The PLCγ pathway generates IP3 and DAG, inducing intracellular calcium release, which drives cytoskeletal reorganization and provides the mechanical force for cell migration. The Vav-Rac pathway directly activates the small GTPase Rac, which orchestrates actin polymerization and the formation of lamellipodia, providing the structural basis for cell migration and phagocytosis. Subsequently, microglia migrate directionally along the Aβ concentration gradient toward the plaques. They extend lamellipodia to engulf Aβ, forming phagosomes. These phagosomes then fuse with lysosomes, leading to the degradation of Aβ. Furthermore, TREM2 signaling can suppress excessive activation of the NF-κB pathway via mechanisms involving PI3K-Akt, thereby preventing harmful neuroinflammation. The core mechanism is illustrated in Figure 4. If TREM2 function is impaired (e.g., due to genetic variants), microglia fail to be properly activated, migrate, and clear Aβ, resulting in a loss of their neuroprotective function and ultimately exacerbating AD pathology.

Despite substantial evidence linking inflammation to Alzheimer's disease (AD) pathology, limited longitudinal data and the paucity of very-early biomarkers hinder precise delineation of causal directionality and temporal onset.[165] One model posits that abnormal aggregation of Aβ oligomers or tau initially activates glial cells, precipitating an inflammatory response. An alternative "inflammation-first" hypothesis proposes that early dysregulation of innate immunity precedes and promotes the emergence and spread of Aβ and tau pathologies.[164] These two perspectives may both reflect a complex bidirectional interplay between inflammation and core pathologies throughout the protracted disease course.[174] A pivotal shift is now underway, with recent investigations concentrating on the collapse of communication mechanisms between brain cells. It is proposed that the breakdown in neuron-glia crosstalk, especially involving astrocytes and microglia, may underpin disease advancement.[175]

Proteomic studies have identified novel and major facilitator such as AHNAK(Adductin-H-N-A-K), underscoring a paradigm shift in AD research. This research marks a move beyond a narrow focus on Aβ and tau toward a "systems biology" perspective that seeks to deconstruct the entire dysregulated network of the disease. We are no longer satisfied with hunting for a single "culprit" but are now committed to dismantling the entire "criminal network." [175] There is also research shows that, the implementation of strategies targeting CD44 and CD33, enhancing the expression of HLA-DR, P2RY12, and ApoE, and modulating microenvironmental signals may contribute to slowing or reversing the pathological progression of Alzheimer's disease, thereby providing a scientific rationale for developing novel therapeutics.[176]

In parallel, metabolic influences are increasingly recognized as central to AD pathogenesis. This occurs through a two-pronged pathway: pathological proteins like Aβ and tau trigger metabolic reprogramming that impairs the immune function of microglia, while pre-existing metabolic disturbances may predispose microglia to a dysfunctional state that accelerates disease. This interplay likely differs between sporadic and familial AD; the latter is primarily driven by early Aβ accumulation, whereas the former appears more influenced by systemic metabolic dysfunction and neuroinflammation. Furthermore, metabolic diseases such as obesity and type 2 diabetes, which are strongly associated with an increased risk of AD, may influence brain immunity and microglial function through their associated metabolic hormones, such as insulin, leptin, adiponectin, and glucagon-like peptide-1 (GLP-1). In individuals with these conditions, microglia may exhibit reduced tolerance to pathological stimuli like Aβ, compromising their ability to mount an effective response. This diminished adaptability can lead to defective immune responses and impaired Aβ clearance, ultimately accelerating AD pathogenesis. Looking forward, a key research direction is to elucidate the mechanistic links between metabolic pathways and microglial dysfunction and to identify novel therapeutic targets that can restore microglial homeostasis.[177]

Given its complex dual roles, targeting neuroinflammation presents a unique therapeutic challenge: how to suppress its chronic, destructive outputs while preserving its acute, protective functions. This complexity explains the failure of broad anti-inflammatory drugs (e.g., NSAIDs) in AD trials. The emerging goal is not wholesale suppression, but precision immunomodulation. Strategies now aim to reprogram microglial states—for instance, by enhancing protective pathways like TREM2 signaling to boost Aβ clearance, or by targeting specific inflammatory nodes (e.g., NLRP3, CD33) to curb cytokine production without compromising phagocytosis. The functional state of microglia is thus a critical determinant of therapeutic efficacy, influencing even the response to anti-Aβ immunotherapies.

This nuanced view is part of a broader paradigm shift towards a "systems biology" perspective of AD. Research is moving beyond a narrow focus on individual pathologies to deconstruct the entire dysregulated cellular network. This includes understanding how metabolic disturbances (e.g., in obesity or diabetes) can predispose microglia to a dysfunctional, pro-inflammatory state, thereby lowering the brain's resilience to Aβ and tau. This systems-level understanding underscores that successful therapeutic intervention will require a multi-target approach. It necessitates combining Aβ- and tau-directed therapies with immunomodulatory agents that are precisely timed and tailored to an individual's neuroinflammatory status.

In conclusion, neuroinflammation is a central pathogenic amplifier in AD, inextricably linking Aβ and tau pathologies into a coordinated destructive network. Its presence underscores the limitation of any single-pathway intervention. Taming this maladaptive immune response requires early, biomarker-guided detection of neuroinflammatory states and combination regimens that include precision immunomodulation, alongside amyloid and tau therapeutics, to effectively disrupt the disease cycle.

**3.2.2 Interplay among Aβ, Tau and Neuroinflammation**

As outlined in the previous section, neuroinflammation is intricately linked with both Aβ and tau pathologies. Building upon that foundational understanding, this section will delve deeper into the mechanistic interaction of pathways of Aβ, tau and neuroinflammation.

While neuroinflammation initially represents a protective response to central nervous system (CNS) insults, its chronic activation in Alzheimer's disease (AD) profoundly exacerbates core pathological processes, establishing a self-perpetuating cycle of neurodegeneration.[163] This maladaptive immune response not only fails to resolve underlying pathologies but actively amplifies them through multiple interconnected mechanisms. In this process, Aβ and tau act as principal pathological drivers that potently induce the activation of astrocytes and microglia,[110] which in turn release a spectrum of pro-inflammatory cytokines such as tumor necrosis factor alpha (TNF-α) and interleukin-1β (IL-1β), along with reactive oxygen species (ROS) and reactive nitrogen species (RNS),[178] thereby triggering neuroinflammation.[179] However, this inflammatory response has a dual role: while it may provide protective effects by enhancing Aβ degradation and clearance,[180] it can also lead to excessive production of Aβ and tau proteins,[181,182] ultimately inducing neurodegeneration and synaptic loss.

In addition to activating astrocytes and microglia, Aβ and tau proteins also exacerbate each other's pathological effects, thereby intensifying neuroinflammation. Pathological tau protein disrupts Aβ clearance mechanisms by at least two ways, leading to persistent Aβ accumulation. One way is that pathological tau protein activates microglia and astrocytes, sustaining a chronic inflammatory state and impairing microglia's capacity to engulf and clear Aβ.[183] Aβ progressively accumulates, and this accumulated Aβ sustains inflammatory responses by activating microglial NLRP3 inflammasomes and other pathways. The other way is that tau-laden neurons release factors that suppress microglial phagocytosis and reduce expression of Aβ-degrading enzymes such as neprilysin.[184,185] Furthermore, it has been proposed that extracellular tau species might promote Aβ aggregation through mechanisms such as cross-seeding or by stabilizing toxic Aβ oligomers.[183,186] The newly formed, structurally more stable Aβ aggregates (particularly oligomers and plaques) serve as potent activation signals for glial cells.[187] This enhanced stimulation promotes excessive activation of microglia and astrocytes, triggering inflammatory responses that release excessive pro-inflammatory cytokines and reactive oxygen species, thereby significantly exacerbating overall neuroinflammation.[188]

Neuroinflammation directly accelerates tau pathology by activating a network of kinases that drive tau hyperphosphorylation. Pro-inflammatory cytokines, particularly TNF-α and IL-1β, stimulate signaling pathways that activate several tau kinases, most notably glycogen synthase kinase-3β (GSK-3β)[189,190] and p38 mitogen-activated protein kinase (p38 MAPK).[191] For instance, TNF-α can activate GSK-3β via the PKCδ pathway, leading to phosphorylation of tau at pathological epitopes such as AT8 (Ser202/Thr205)[192] and PHF-1 (Ser396/404).[193] Moreover, inflammasome-derived IL-1β has been shown to enhance tau phosphorylation both in vitro and in vivo,[155,194] contributing to the formation of neurofibrillary tangles (NFTs). Critically, during the pathological progression of AD, Aβ and tau proteins may trigger neuroinflammation, which in turn acts as a partial driver of tau pathology, collectively forming an intricate, destructive pathological cycle. Conversely, tau pathology indirectly promotes neuroinflammation by impeding Aβ clearance,[172] creating another positive feedback loop. Pathological tau protein disrupts Aβ clearance mechanisms by at least two ways, leading to persistent Aβ accumulation. One way is that pathological tau protein activates microglia and astrocytes, sustaining a chronic inflammatory state and impairing microglia's capacity to engulf and clear Aβ.[183] This directly results in the persistent accumulation of Aβ, which subsequently drives and escalates inflammatory responses through mechanisms such as activating NLRP3 inflammasomes in microglia. The other way is that tau-laden neurons release factors that suppress microglial phagocytosis and reduce expression of Aβ-degrading enzymes such as neprilysin.[184,185] Furthermore, it has been proposed that extracellular tau species might promote Aβ aggregation through mechanisms such as cross-seeding or by stabilizing toxic Aβ oligomers.[183,186] The newly formed, structurally more stable Aβ aggregates (particularly oligomers and plaques) serve as potent activation signals for glial cells.[187] This enhanced stimulation promotes excessive activation of microglia and astrocytes, triggering inflammatory responses that release excessive pro-inflammatory cytokines and reactive oxygen species, thereby significantly exacerbating overall neuroinflammation.[188] At this point, Aβ, tau, and neuroinflammation form an inescapable destructive cycle, culminating in irreversible structural damage to brain networks, such as the abnormal clearance of synapses primarily mediated by the complement system. In the AD brain, chronic inflammatory conditions upregulate complement components such as C1q and C3.[195] These proteins tag synapses for elimination by engaging complement receptors (e.g., CR3) on microglia,[196] triggering excessive, activity-independent phagocytosis of synaptic structures[197]—a process analogous to developmental synaptic pruning but pathologically activated in the adult brain.[198] This complement-mediated synaptic stripping is further amplified by Aβ oligomers,[199] which can directly activate the classical complement pathway.[200] The resultant loss of excitatory synapses, particularly in the hippocampus and cortex,[201] is a major structural correlate of cognitive decline in AD. Thus, neuroinflammation transforms a physiological pruning mechanism into a destructive process that directly undermines neural circuit integrity and cognitive function.

In summary, neuroinflammation acts as a central pathological amplifier in AD, impairing Aβ clearance, promoting tau hyperphosphorylation, and driving complement-mediated synaptic loss. These mechanisms collectively underscore that inflammation is not a passive bystander but an active driver of disease progression. Targeting specific inflammatory pathways—while preserving beneficial immune functions—thus represents a critical therapeutic strategy for disrupting this destructive cascade.

### 3.3. Cholinergic hypothesis

The cholinergic hypothesis, which posits a central role for acetylcholine (ACh) deficiency in Alzheimer's disease (AD), has historically provided the foundation for symptomatic pharmacotherapy. Acetylcholine (ACh), the first neurotransmitter identified in humans, is integral to signal transmission in both the central and peripheral nervous systems.[202] Basal forebrain cholinergic projections innervate the neocortex and hippocampus—regions essential for learning, memory, and higher-order cognition.[203]. ACh is synthesized in cholinergic neurons and is crucial for synaptic plasticity, attention, and memory consolidation.[83] In AD, the degeneration of basal forebrain cholinergic neurons and the consequent reduction in cortical ACh levels are well-established[203] and contribute significantly to cognitive decline.[63,204-206] Consequently, degeneration of basal forebrain cholinergic neurons lowers cortical ACh levels,[207] contributing to memory impairment and broader cognitive deficits in AD.[66,208-210]

Acetylcholinesterase, located within the synaptic cleft and on the postsynaptic membrane, rapidly hydrolyzes ACh, terminating synaptic signaling.[118] Choline liberated by this reaction is recaptured by presynaptic high-affinity choline transporters (CHT) and reutilized for ACh resynthesis, with a recycling efficiency exceeding 50%.[211,212]

The clinical translation of this hypothesis led to the development of cholinesterase inhibitors, such as donepezil and galantamine—augment ACh neurotransmission by inhibiting its enzymatic degradation. While these agents have demonstrated modest benefits on cognition, activities of daily living, and can stabilize symptoms or slow clinical deterioration over approximately 6–12 months,[213-215] their inability to halt disease progression underscores a critical limitation: targeting a single downstream neurotransmitter system is insufficient against a multifactorial pathology.

This limitation of monotherapy informs the first paradigm shift. Cholinergic deficits themselves may serve as valuable biological markers. Imaging or biochemical assays detecting cholinergic neuron loss or functional decline could be integrated into a biomarker-guided framework for earlier and more precise diagnosis, before widespread cognitive impairment occurs. Given the multisystem nature of Alzheimer's disease (AD), the proposition that cholinergic deficiency constitutes the initiating lesion remains contested.[204]

Furthermore, the cholinergic system does not exist in isolation but interacts extensively with core AD pathologies. For instance, Antimuscarinic agents, widely used in anesthetic practice, attenuate muscarinic receptor–mediated ACh signaling;[212,216,217] however, their therapeutic value in AD remains uncertain and may be limited by adverse cognitive effects.[218] The cholinergic system comprises multiple receptor subtypes with distinct roles and alterations in AD, including muscarinic receptors (M1–M5) and nicotinic receptors (e.g., α4β2, α7). M1 receptors are closely linked to cognitive processing,[219] whereas the α7 nicotinic receptor has been implicated in Aβ interactions. These observations suggest that future therapeutics may require more precise targeting of specific receptor subtypes to optimize efficacy and minimize side effects.[220] This intricate interplay provides the scientific rationale for the second paradigm shift: the necessity of combination therapies. The future of AD management lies not in abandoning cholinergic modulation, but in strategically integrating it with interventions that target the core disease drivers, such as anti-amyloid-β and anti-tau agents, as well as anti-inflammatory strategies. It offers a rational multi-target approach that may enhance clinical efficacy and mitigate the limitations of monotherapy.[221]

In conclusion, the legacy of the cholinergic hypothesis is not merely the symptomatic treatments it inspired, but the profound lesson it teaches. It compellingly demonstrates that effective AD management must evolve beyond singular, late-stage interventions toward a new model defined by preemptive, biomarker-driven diagnosis and multi-target, disease-modifying therapeutic combinations.

### 3.4 Glutamatergic excitotoxicity

Glutamate, the central nervous system's primary excitatory neurotransmitter,[222] plays a dual role in Alzheimer's disease (AD), underpinning both normal cognition and a key pathological process known as excitotoxicity.[223,224] Under physiological conditions, its signaling is tightly regulated. However, in the AD brain, multiple upstream pathologies converge to disrupt this balance, inadequate glutamine supply impairs neuronal glutamate synthesis, prompting compensatory increases in presynaptic glutamate release to sustain neurotransmission.[223,225,226] Upon arrival of an action potential at the presynaptic terminal, opening of voltage-gated calcium channels (VGCCs) permits Ca2+ influx, triggering synaptic vesicle fusion and exocytotic release of glutamate.[227,228] Pathological dysregulation can lead to uncontrolled glutamate release, elevating synaptic cleft concentrations from physiological micromolar to neurotoxic millimolar levels.[229-231]

Core AD pathologies directly fuel this excitotoxic cascade. Aβ oligomers can impair astrocytic glutamate reuptake by EAAT1/2 transporters and sensitize postsynaptic NMDA receptors, while tau pathology and neuroinflammation further disrupt neuronal and glial homeostasis. This confluence of factors leads to a vicious cycle: impaired glutamate clearance prolongs its synaptic presence, causing excessive activation of NMDA and AMPA receptors. The resultant massive $Ca^{2+}$ influx[231] into neurons initiates cytotoxic cascades, Therapeutically, validation of the glutamate excitotoxicity model supports neurotransmitter modulation as a means to ameliorate cognitive symptoms. Cold Spring Harbor perspectives in medicineaccumulation of Aβ plaques and tau neurofibrillary tangles promotes excess glutamate release, resulting in pathological overactivation of NMDA receptors and excessive Ca2+ influx, which culminates in excitotoxic neuronal death. Therapeutically, an optimal NMDA receptor antagonist for AD should satisfy three principal criteria: preferential blockade under conditions of pathological overactivation (i.e., voltage dependence), relatively low binding affinity to avoid prolonged tonic inhibition, and sufficient lipophilicity and pharmacokinetics to cross the blood–brain barrier and act centrally.[232]

Memantine's tricyclic, hydrophobic adamantane-based scaffold confers voltage-dependent channel block, enabling preferential targeting of overactivated NMDA receptors. The primary amine at the 1-position facilitates protonation at physiological pH, supporting selective electrostatic interactions within the channel pore that enhance specificity while maintaining appropriately weak affinity.[233] This profile allows effective attenuation of pathological Ca2+ influx (neuroprotection) with rapid unbinding under physiological activity, thereby preserving normal synaptic function.[68] Additionally, methyl substituents at the 3,5-positions increase lipophilicity, promoting blood–brain barrier penetration and adequate CNS exposure.[234] Collectively, these properties underlie memantine's status as the only NMDA receptor antagonist approved for long-term treatment of AD.

Despite their symptomatic benefits, acetylcholinesterase inhibitors (AChEIs) and NMDA receptor antagonists primarily modulate neurotransmission and do not modify the underlying disease trajectory.[235,236] This fundamental limitation not only defines the ceiling of traditional care but also clearly dictates the imperative for the first paradigm shift: the pursuit of disease-modifying therapies (DMTs) that target core pathologies.[237] Disease-Modifying

Therapy (DMT) is defined as a treatment strategy designed to target the underlying etiology or core pathological pathways of a disease itself, aiming to delay, halt, or potentially reverse the disease progression, in contrast to the "symptomatic treatment" provided by conventional drugs. Two phase IV clinical trials are ongoing through 2025 to further evaluate efficacy and safety.[238]

## 3.5 An Integrated Network Model: From Linear Hypotheses to a Pathological Cascade

The preceding sections have detailed the principal pathological pathways in Alzheimer's disease (AD) as distinct entities. However, the repeated failure of therapies targeting single pathways underscores a fundamental insight: AD would rather be a syndrome driven by a dynamic, self-reinforcing pathological network than a disorder of a linear cascade.[174,239] This network view explains why the disease, once initiated, gains momentum and becomes progressively independent of its initial trigger. Converging proposal posit that neuroinflammation interplays with Aβ and tau synergistic effects, seeming to be a pivotal factor in AD's pathophysiology.[108,240]

The disease cascade is often catalyzed by an initial disruption in Aβ homeostasis, primarily impaired clearance in sporadic AD. The resulting accumulation of Aβ assemblies, particularly soluble oligomers, acts as a chronic "danger signal." As detailed in Section 3.1.1, these oligomers bind neuronal receptors, triggering intracellular calcium dysregulation and oxidative stress. This compromised neuronal environment is the first critical step in creating a permissive milieu for the amplification of downstream pathologies. It is important to note that at this stage, significant cognitive symptoms may be absent, representing the long preclinical phase of AD.[241]

The compromised neuronal environment and the presence of Aβ plaques potently activate microglia, primarily via receptors like TREM2 (Section 3.2.1). Initially, this is a protective response aimed at clearing debris. However, sustained activation leads to a maladaptive state characterized by two key failures: a diminished capacity for Aβ phagocytosis and an exaggerated release of pro-inflammatory cytokines (e.g., IL-1β, TNF-α).[242] This inflammatory milieu, as explored in Section 3.2.2, directly fuels the tau pathology cascade. For instance, TNF-α signaling can activate kinases like GSK-3β and p38 MAPK, which hyperphosphorylate tau, facilitating its dissociation from microtubules and misfolding.

This inflammation-driven hyperphosphorylation marks the critical transition of tau from a passive bystander to an active executing agent. As described in Section 3.1.2, pathological tau gains prion-like properties, enabling its trans-synaptic spread throughout brain networks.[243] This spread of tau pathology (Braak staging) correlates strongly with the emergence and progression of clinical symptoms. Furthermore, tau itself becomes a novel inflammatory stimulus, further activating glial cells and solidifying the chronic inflammatory state. This creates a destructive cycle, wherein tau-driven inflammation further perturbs neuronal and glial homeostasis, indirectly promoting continued Aβ accumulation.

While the synergy of Aβ-Tau-Inflammation drives disease progression, the catastrophic failure of synaptic integrity and plasticity—the direct structural correlate of cognitive decline—unfolds through a coordinated, multi-stage assault. This process constitutes a self-reinforcing synaptic failure loop. The initial insult involves structural and functional sabotage:[244] Aβ oligomers directly bind to and corrupt postsynaptic densities, while pathological tau disrupts axonal transport, starving synapses of essential components. These initial injuries are dramatically amplified by secondary excitotoxicity and active elimination. The corrupted synaptic environment leads to pathological NMDA receptor overactivation (Section 3.4.2), while the inflamed milieu activates the complement cascade, tagging compromised synapses for microglia-mediated phagocytosis.[245] This aberrant pruning erases neural connections, directly underlying the patient's progressive cognitive decline.

This integrated network model provides a compelling explanation for clinical observations and therapeutic failures. It clarifies why targeting Aβ alone in symptomatic patients fails: by that stage, the tau and neuroinflammation pathways have become self-sustaining drivers of neurodegeneration. Similarly, broadly suppressing inflammation may blunt protective functions, and intervening against tau after its widespread dissemination is likely "too late".

Therefore, the therapeutic imperative is unequivocal: successful disease modification will require multi-target, combination strategies that are initiated early, based on biomarker profiles.[246,247] The model dictates that the optimal window for intervention is before the system reaches criticality—that is, during the preclinical or prodromal stages when Aβ pathology is present, but tau pathology and inflammation are not yet widespread and self-sustaining.[248]

This network perspective directly informs the next generation of clinical trials and the diagnostic paradigm shift discussed in Chapter 4. For example, a rational combination regimen might include: an anti-Aβ monoclonal antibody (e.g., lecanemab) to reduce the initial driver; a tau aggregation inhibitor to halt the spread of tangles; and a precision immunomodulator (e.g., a TREM2 agonist) to enhance protective microglial functions. The future of AD management lies in deconstructing this pathogenic network through preemptive, biomarker-guided application of such personalized, multi-pronged therapeutic assaults.

**4.Diagnosis:The New Paradigm in AD Management**

The effective management of Alzheimer's disease is undergoing a fundamental transformation, pivoting on two interdependent paradigm shifts: biomarker-guided early diagnosis and combination therapies targeting multifactorial pathology. The validation and standardization of core biomarkers are the essential enablers of this new strategic approach.

**The AT(N) Framework: A Biological Blueprint for Precision Medicine**

The National Institute on Aging and the Alzheimer's Association (NIA–AA) have proposed a research framework that characterizes the biology of Alzheimer's disease (AD) using the AT(N) system—amyloid-β deposition (A), pathologic tau (T), and neurodegeneration (N) (**Figure 5**) —as core biomarker categories.[249] Established biomarkers principally encompass neuroimaging measures, cerebrospinal fluid (CSF) analytes, and blood-based assays. These biomarkers are commonly assessed with magnetic resonance imaging (MRI) and positron emission tomography (PET), which enable in vivo characterization of structural and functional brain alterations.[38,250]

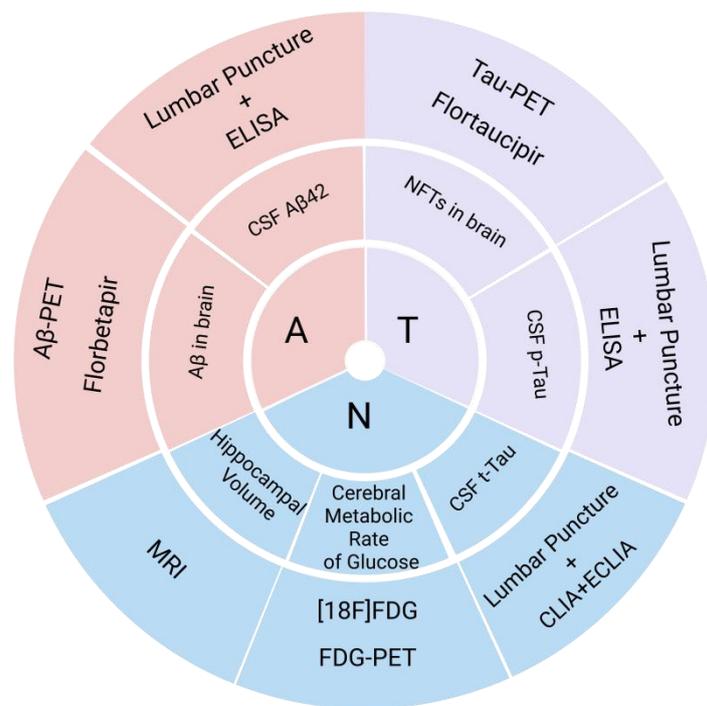

**Figure 5** A-T-N Biomarker Scheme in Alzheimer's Disease: A Conceptual Overview

In the context of neuroimaging for AD, the principal aims are early detection of characteristic pathophysiological changes, exclusion of alternative etiologies of cognitive impairment, and longitudinal evaluation of disease progression.[36,251] Each imaging modality confers distinct strengths. MRI affords high-resolution delineation of gray and white matter and is particularly sensitive to atrophic changes in the hippocampus and entorhinal cortex—regions affected early in the course of AD.[251] Relative to cognitively normal older adults, individuals with AD demonstrate accelerated hippocampal atrophy.[252] Accordingly, volumetric MRI of the hippocampus can assist in identifying AD-specific morphometric alterations.[253] In subsequent clinical management, MRI is frequently employed for serial quantification of hippocampal volume to monitor disease trajectory and inform therapeutic decision-making. The specific value in early diagnosis lies in the fact that although hippocampal atrophy is A manifestation of neuronal loss (i.e., neurodegeneration (N)) and appears relatively later than upstream pathologies such as Aβ (A), it remains an important marker of the transition from the preclinical stage to the symptomatic stage. In individuals with mild subjective cognitive decline, the detection of definite medial temporal lobe atrophy can significantly increase the diagnostic confidence of AD-derived mild cognitive impairment (MCI), thereby accurately identifying this high-risk group from numerous cognitive complainants and achieving pre-symptom or very early diagnostic intervention. In clinical trials of combination therapy and future clinical practice, the brain structural changes revealed by MRI (such as the slowdown in atrophy rate) are one of the core biological endpoints for evaluating whether the treatment has a protective effect on the neurodegenerative (N) process. If a certain combination therapy can effectively stabilize the volume of the hippocampus, it proves that it may truly have the disease-modifying effect of delaying the progression of the disease.

Computed tomography (CT) produces cross-sectional structural images of the brain using X-ray–based techniques. It is useful for the initial evaluation of global cerebral atrophy.[251] However, owing to its limited soft-tissue contrast, CT lacks sensitivity for early Alzheimer's disease (AD)–specific abnormalities, including hippocampal atrophy and medial temporal lobe pathology.[253] Consequently, CT is best suited for the rapid exclusion of acute intracranial hemorrhage, large territorial infarction, and other emergent neurological conditions, and it is not recommended for early AD screening.[254] The efficient exclusion of other diseases through CT can ensure that patients who are subsequently included in the "early diagnosis of AD" process and combination therapy clinical trials are more likely to have cognitive impairment due to the neurodegenerative pathology of AD itself rather than other structural lesions. This enhances the purity of clinical diagnosis and trial populations, serving as the foundation for obtaining reliable research results and verifying the effectiveness of combination therapies.

Alzheimer's disease is defined by characteristic, imageable neuropathological hallmarks.[249] Positron emission tomography (PET) exploits these features through radiotracers that bind selectively to the relevant pathological substrates,[255] enabling in vivo visualization via detection of tracer distribution.[256] A central feature of AD pathophysiology is β-amyloid (Aβ) deposition,[257] which typically occurs early—often 5 to 20 years before symptom onset[250,258]—and contributes to synaptic dysfunction through the formation of senile plaques.[259] Aβ-targeted PET tracers are engineered to cross the blood–brain barrier and bind specifically to fibrillar Aβ.[260] After intravenous administration, these tracers accumulate in regions with high Aβ burden.[260] The PET scanner detects positron emissions to generate quantitative or semi-quantitative maps of cerebral Aβ distribution, with signal intensity reflecting plaque density.[261] In clinical practice, 18F-labeled tracers are preferred due to their favorable half-life and logistics, with commonly used agents including [18F]flutemetamol,[262] [18F]florbetapir,[261] and [18F]florbetaben.[263] Because Aβ deposition represents an early event in the AD cascade, Aβ PET can reveal pathological changes prior to overt cognitive decline and thereby assists in distinguishing normal aging from preclinical AD.[264] The core value of Aβ PET in early diagnosis lies in its ability to provide an objective biological diagnosis for "preclinical AD", significantly advancing the intervention window before symptoms appear. At the same time, in the decision-making of combination therapy, Aβ positive (A+) confirmed by Aβ PET is the cornerstone of individualized treatment plans and A prerequisite for determining whether patients can benefit from anti-Aβ therapy, thereby providing a key basis for precisely combining therapies targeting other pathologies such as Tau or neuroinflammation.

Tau-targeted radiotracers, analogous in concept to Aβ ligands, are engineered to bind selectively to hyperphosphorylated tau[265]—the principal constituent of neurofibrillary tangles (NFTs)—with minimal affinity for physiologic tau isoforms.[266] PET imaging detects emitted radioactivity to produce spatial and quantitative maps of NFT burden,[267] wherein signal intensity reflects the extent of neuronal injury and loss.[127,265] NFTs constitute a proximate neurotoxic pathology in Alzheimer's disease (AD);[101] their accumulation, driven by aggregated hyperphosphorylated tau, is strongly associated with cognitive decline, and higher NFT burdens correlate with greater clinical impairment. Consequently, tau PET tracers have substantial diagnostic[267] and prognostic utility.[127,128]

Although tau pathology occurs across multiple neurodegenerative syndromes, disease-specific topographies are informative.[268] In AD, NFTs predominantly involve temporal and parietal cortices, whereas in frontotemporal lobar degeneration (FTLD) they more commonly localize to frontal regions.[14] This regional specificity facilitates differential diagnosis among dementia subtypes.[127] Moreover, NFT propagation in AD follows a stereotyped trajectory, typically beginning in the medial temporal lobe and extending to parietal and then frontal cortices. Tau PET can therefore support disease staging: confinement to medial temporal structures suggests an early stage, while widespread neocortical involvement indicates moderate to advanced disease.[269] The most extensively validated tau tracer in current clinical use is

[18F]flortaucipir (T807). Next-generation ligands are under active development to improve specificity, off-target binding profiles, and quantitative performance.[267] The unique value of Tau PET in early diagnosis lies in that its medial temporal lobe deposition may even occur earlier than detectable Aβ pathology, providing another key window for identifying AD-specific lesions at an extremely early stage. In terms of providing A basis for combination therapy, the entanglement load and distribution revealed by Tau PET are the most crucial biological endpoints for evaluating disease aggressiveness, selecting the applicable population for tau-targeted therapy, and directly monitoring its efficacy, thereby guiding the precise combination and application timing of anti-A β therapy and anti-tau therapy.

A third class of PET radiotracers assesses cerebral metabolism rather than binding directly to pathological proteins.[256] These agents index regional glucose utilization,[270] a proxy for synaptic and neuronal activity.[271,272] [18F]FDG, a radiolabeled glucose analog, is transported into neurons and phosphorylated but not further metabolized, resulting in intracellular trapping.[273] PET quantification of [18F]FDG distribution thus reflects regional metabolic demand; reduced signal denotes hypometabolism consistent with neuronal dysfunction or loss. In AD, hypometabolism typically appears in bilateral parietotemporal association cortices—regions crucial for memory and language—emerging after Aβ deposition and with advancing tau pathology.[274,275] Distinct hypometabolic patterns aid differential diagnosis: AD is characterized by relatively symmetric parietotemporal reductions,[276] whereas vascular dementia often shows multifocal, patchy deficits aligned with vascular territories.[277] Accordingly, [18F]FDG-PET contributes to subtype differentiation, disease severity assessment, and therapeutic planning.[278] The key role of [18F]FDG-PET in early diagnosis lies in its ability to sensitively capture the functional dysregulation of the characteristic brain networks of AD. This metabolic abnormality often occurs before structural atrophy becomes widespread, providing an important functional imaging basis for identifying AD-specific pathology in the stage of mild cognitive impairment (MCI). In terms of providing A basis for combination therapy, it serves as an effective indicator of neurodegeneration (N), capable of assessing the overall functional impact of the disease and offering A crucial efficacy monitoring endpoint for measuring whether the treatment has successfully stabilized or improved the metabolic and functional status of the brain, thereby completing the evaluation system of combined treatment strategies based on A (Aβ) and T (Tau)）

The core pathological processes of Alzheimer's disease (AD)—β-amyloid deposition, tau pathology, and neuronal injury—are reflected in cerebrospinal fluid (CSF) and blood.[250] Accordingly, CSF- and blood-based assays constitute key biomarker modalities for early diagnosis, disease monitoring, and differential diagnosis by capturing these molecular signatures.[279] CSF, which bathes the brain and spinal cord, is in direct exchange with the interstitial milieu and thus most faithfully mirrors intracerebral pathology. It remains among the most specific and accurate biomarker sources for AD.[280] Principal CSF biomarkers include the Aβ42/Aβ40 ratio, total tau (t-tau), phosphorylated tau (p-tau), and neurofilament light chain (NfL). In particular, the combination of the Aβ42/Aβ40 ratio with p-tau provides a direct correlate of AD pathology and achieves diagnostic accuracies exceeding 90%.[281] In early diagnosis, the core value of cerebrospinal fluid biomarkers lies in their extremely high sensitivity and specificity, which can objectively confirm the pathological presence of A (Aβ) and T (Tau) in AD several years or even decades before the appearance of clinical symptoms. They are one of the gold standards for achieving "preclinical AD" diagnosis. In terms of providing A basis for combination therapy, the precise A/T/N typing offered by cerebrospinal fluid testing serves as a direct blueprint for formulating individualized combination strategies: For instance, the profile of A+T+N+ strongly suggests the need for A combined therapy targeting both Aβ and tau simultaneously, while continuous monitoring of cerebrospinal fluid p-tau and NfL levels can provide dynamic and quantitative molecular evidence for evaluating the efficacy of anti-tau treatment and neuroprotection.）

Despite the low concentrations of Aβ and p-tau that traverse the blood–brain barrier, ultra-sensitive platforms—such as single-molecule array (Simoa) and digital ELISA—enable detection of these analytes, supporting noninvasive screening.[280] Nonetheless, plasma biomarkers can be affected by peripheral factors (e.g., renal or hepatic dysfunction and other systemic conditions), which may modestly reduce specificity relative to CSF measures; mild p-tau elevations can occasionally occur in non-AD contexts. Moreover, at the very earliest disease stages—when Aβ deposition is nascent and tau pathology limited—blood assays are generally less sensitive than CSF biomarkers. In such scenarios, confirmatory neuroimaging (e.g., PET) may be warranted.[282] However, updated diagnostic platforms are under active development. Recently, the first fully automated immunoassay for brain-derived tau (BD-Tau) in plasma has been introduced. This assay specifically targets BD-Tau, a variant designed to better capture tau proteins of central origin, thereby minimizing confounding signals from peripheral tau. This fundamental distinction sets it apart from conventional manual ELISA or digital immunoassay platforms. The selective targeting of BD-Tau enhances the assay's value in interpreting Alzheimer's disease-specific pathology, underscoring its superior specificity.

## 5.Therapeutic Strategies for Alzheimer's Disease: From Traditional Pharmacotherapy to Cutting-Edge Technologies

### 5.1 Traditional Drugs

Pharmacological therapy remains a cornerstone of Alzheimer's disease (AD) management and, when integrated with emerging technologies, has the potential to substantially advance clinical care. Clinically utilized agents are conventionally divided into two principal classes:

acetylcholinesterase inhibitors (AChEIs)[283]—including tacrine, donepezil, rivastigmine, and galantamine—and N-methyl-D-aspartate (NMDA) receptor antagonists, represented by memantine.[284] By augmenting cholinergic neurotransmission—principally through increasing synaptic acetylcholine availability and prolonging its action[66]—these agents can yield modest improvements in cognition and behavioral symptoms.[285]

Tacrine, the prototypical first-generation AChEI approved in 1993, has largely fallen out of use due to hepatotoxicity and other adverse effects.[286] Second-generation AChEIs (e.g., donepezil, rivastigmine, and galantamine) demonstrate greater target selectivity, **Figure 6** improved tolerability, and more favorable pharmacokinetic properties,[203] and they are now considered first-line symptomatic treatments for AD.[287] Memantine, an uncompetitive NMDA receptor antagonist, is typically employed in moderate to severe stages,[235] either as monotherapy or in combination with AChEIs, to further address excitotoxic mechanisms and support symptomatic benefit.

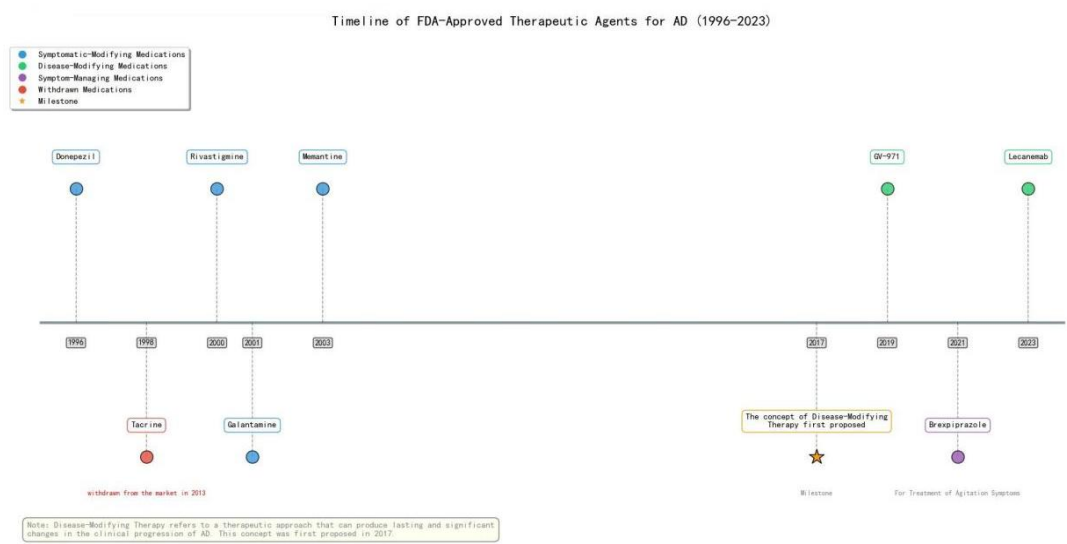

**Figure 6** Alzheimer's Disease Drug Development Timeline: Key Agents and Clinical Translations

The efficacy and broad clinical adoption of these agents derive from their distinctive chemical architectures. Acetylcholinesterase inhibitors (AChEIs) constitute a mechanistic class rather than a single scaffold, encompassing structurally diverse compounds that share the common pharmacological action of inhibiting acetylcholinesterase (AChE).[288] Their design principles typically involve either mimicking essential structural motifs of acetylcholine (ACh), the endogenous AChE substrate, or engaging AChE's active site through specific functional groups to prevent ACh hydrolysis.[289] Across AChEIs, interactions center on the enzyme's anionic and esteratic subsites;[288] however, the core ring systems and

substituents differ substantially among individual drugs.[290]

The first-generation AChEI, tacrine, comprises a tricyclic framework of two benzene rings fused to an acridine ring (a nitrogen-containing heteroaromatic), with a tertiary amine substituent. This configuration confers limited selectivity for central versus peripheral AChE, leading to off-target inhibition and a higher incidence of adverse effects.[291] By contrast, second-generation AChEIs incorporate fundamentally different core scaffolds and functional groups, changes that underpin improved selectivity, pharmacokinetic profiles, and overall clinical tolerability and efficacy.[292]

Donepezil incorporates an indanone-derived moiety linked to a piperidine ring;[293] this scaffold affords high affinity for acetylcholinesterase (AChE) while, through optimized three-dimensional fit, limiting engagement of peripheral AChE isoforms (e.g., cardiac), thereby reducing off-target adverse effects.[294] Its metabolic stability confers a prolonged elimination half-life of approximately 70–80 hours. Rivastigmine employs a carbamate warhead that forms a reversible covalent bond with AChE, enabling sustained inhibition at relatively low concentrations.[295] Its compact, hydrophobic indane-like structure is associated with reduced hepatotoxic liability and avoids the tricyclic metabolic issues observed with first-generation tacrine.[283] Galantamine, a rigid, naturally derived alkaloid, exhibits favorable metabolic stability;[296] allylamine substitution attenuates peripheral cholinergic effects and supports suitability for long-term administration.[297]

In clinical practice, combination strategies may enhance outcomes.[298] Evidence suggests that using appropriate cholinesterase inhibitors in combination (e.g., donepezil with galantamine) or pairing an AChEI with other neuroactive agents, metal chelators, or antioxidants can, in selected contexts, improve efficacy, tolerability, and safety profiles in AD management;[299] such approaches should be individualized and evidence-informed.

The second major class of clinically validated agents comprises N-methyl-D-aspartate (NMDA) receptor antagonists, represented by memantine. Approved by the U.S. Food and Drug Administration for moderate to severe AD, memantine acts as an uncompetitive NMDA receptor blocker, modulating glutamatergic neurotransmission and, indirectly, dopaminergic pathways. Clinically, it can improve cognitive performance, activities of daily living, and behavioral symptoms.

Throughout the course of Alzheimer's disease (AD), accumulation of Aβ plaques and tau neurofibrillary tangles promotes excess glutamate release, resulting in pathological overactivation of NMDA receptors and excessive Ca2+ influx, which culminates in excitotoxic neuronal death. Therapeutically, an optimal NMDA receptor antagonist for AD should satisfy three principal criteria: preferential blockade under conditions of pathological overactivation (i.e., voltage dependence), relatively low binding affinity to avoid prolonged tonic inhibition,

and sufficient lipophilicity and pharmacokinetics to cross the blood–brain barrier and act centrally.[68]

Memantine's tricyclic, hydrophobic adamantane-based scaffold confers voltage-dependent channel block, enabling preferential targeting of overactivated NMDA receptors. The primary amine at the 1-position facilitates protonation at physiological pH, supporting selective electrostatic interactions within the channel pore that enhance specificity while maintaining appropriately weak affinity.[233] This profile allows effective attenuation of pathological $Ca^{2+}$ influx (neuroprotection) with rapid unbinding under physiological activity, thereby preserving normal synaptic function.[68] Additionally, methyl substituents at the 3,5-positions increase lipophilicity, promoting blood–brain barrier penetration and adequate CNS exposure.[234] Collectively, these properties underlie memantine's status as the only NMDA receptor antagonist approved for long-term treatment of AD.

Despite their symptomatic benefits, acetylcholinesterase inhibitors (AChEIs) and NMDA receptor antagonists primarily modulate neurotransmission and do not modify the underlying disease trajectory.[235,236] This fundamental limitation not only defines the ceiling of traditional care but also clearly dictates the imperative for the first paradigm shift: the pursuit of disease-modifying therapies (DMTs) that target core pathologies.[237] Disease-Modifying Therapy (DMT) is defined as a treatment strategy designed to target the underlying etiology or core pathological pathways of a disease itself, aiming to delay, halt, or potentially reverse the disease progression, in contrast to the "symptomatic treatment" provided by conventional drugs. Two phase IV clinical trials are ongoing through 2025 to further evaluate efficacy and safety.

**5.2 The First Paradigm Shift: Monoclonal Antibodies and the Disease Modification**

Monoclonal antibodies targeting Aβ—aducanumab,[300,301] lecanemab,[95,302] and donanemab[96]—represent a major therapeutic advance. Aducanumab binds Aβ residues 3–7 and targets both soluble oligomers and insoluble fibrils;[94,303] lecanemab exhibits high affinity for soluble Aβ aggregates and was developed against the E22G (Arctic) mutation epitope;[304] donanemab preferentially recognizes deposited plaque structures.[303] Collectively, these agents diversify therapeutic targets and deepen mechanistic interrogation of AD, with the potential to slow or halt disease progression.

Although clinically approved drugs have demonstrated certain efficacy, conventional pharmacotherapy continues to face two major challenges: limited therapeutic impact on Alzheimer's Disease (AD) and a high failure rate in drug development. Comprehensive evaluations indicate that current symptomatic treatments can, at best, delay only approximately one-third of the cognitive decline. Furthermore, they are unable to reverse fundamental structural damage, such as hippocampal atrophy. The majority of their benefits are confined to the early and middle stages, leaving a significant therapeutic void for late-stage disease. Moreover, the field has witnessed frequent failures in recent AD drug development, a predicament stemming from the difficulty in balancing the "lack of treatment options" against "medication-related risks." However, the limitations of Aβ-targeting antibodies—their inability to halt cognitive decline completely and the risk of adverse events such as amyloid-related imaging abnormalities (ARIA)—clearly demonstrate that targeting Aβ pathology alone is insufficient to arrest the entire disease process. This critical insight serves as the compelling impetus for the next evolutionary step: the shift towards combination therapies that address the multifactorial nature of AD.

**5.3 Targeting the Multifactorial Pathological Network**

The demonstrated limitations of monotherapy underscore the biological rationale for combination approaches. Given that AD is driven by the complex interplay of amyloidosis, tauopathy, neuroinflammation, and vascular dysfunction, effective disease modification likely requires concurrently addressing multiple pathological axes. This section explores promising therapeutic avenues beyond Aβ, whose integration with existing agents forms the cornerstone of the next-generation combinatorial strategy.

A primary research focus has been on therapeutic pathways derived from different hypotheses, which have demonstrated preliminary efficacy, particularly concerning pathologies involving the tau and neuroinflammatory pathways. The abnormal aggregation of tau protein demonstrates a stronger correlation with the cognitive symptoms and severity of Alzheimer's disease than Aβ. Consequently, reducing neurofibrillary tangles (NFTs) formed by tau in the brains of Alzheimer's patients represents a promising strategy to halt disease progression. However, these fibrils are exceptionally stable, and no drug or method has yet been able to effectively degrade them in the brain. Recent research, however, has revealed that D-peptides can self-assemble into amyloid-like fibers that accumulate on the surface of tau fibrils. This interface leads to fibril fragmentation through a stress-release mechanism, thereby achieving the depolymerization of the ultra-stable tau filaments.[305] Previous studies by the same research team have confirmed that D-peptides of the same series (D-TLKIVWC) can ameliorate behavioral deficits in a mouse model of Alzheimer's disease.[306,307] The progression of these tau-targeting and anti-inflammatory agents into clinical trials will heavily rely on biomarker-guided patient selection (e.g., Tau-PET positivity or specific neuroinflammatory CSF profiles) to identify individuals most likely to benefit, thereby operationalizing the paradigm of precision combination therapy. The development of therapeutics for these diverse targets further underscores the indispensability of the AT(N) biomarker framework. In the future, determining the optimal combination therapy for an individual patient will necessarily depend on their specific biomarker profile—for instance, the relative burdens of Aβ, tau, and neuroinflammation—to ensure the right therapeutic components are matched to the dominant drivers of that individual's disease.

Notably, this entire process requires no enzymatic activity or external energy source. This breakthrough reinvigorates the quest for effective Alzheimer's therapeutics and provides a novel conceptual framework for tackling other amyloid aggregation-related diseases, such as Parkinson's and Huntington's disease, holding the potential to revolutionize the treatment landscape for neurodegenerative disorders.

The treatment strategies informed by the neuroinflammation hypothesis have also witnessed advances in precision medicine, enabling more targeted interventions. Therapeutic avenues reflecting this precision include modulation of TREM2 signaling with agonists (e.g., AL002) to augment tau clearance and temper inflammatory responses,[308] and inhibition of the NLRP3 inflammasome to reduce IL-1β maturation.[309] Agents such as MCC950, by dampening NLRP3 activity, may mitigate tau phosphorylation cascades and limit propagation of tau pathology.[310] The efficacy of anti-inflammatory interventions remains uncertain.[92,93] While epidemiological studies have associated long-term nonsteroidal anti-inflammatory drug (NSAID) use with reduced AD risk,[311] large randomized prevention and treatment trials (e.g., naproxen, rofecoxib, celecoxib) have largely failed to demonstrate cognitive benefit in patients or high-risk cohorts and, in some cases, have increased adverse events.[65,312] These outcomes may reflect suboptimal trial design—such as late intervention, drug selection, and population heterogeneity—but they also challenge a simplistic strategy of broad "inflammation suppression."[313] Consequently, contemporary development efforts prioritize optimal timing and enhanced target specificity.[163,314] Ongoing clinical programs are evaluating agents directed at inflammation-related receptors, signaling pathways, and pro-inflammatory cytokines,[315] with the goal of modulating, rather than indiscriminately suppressing, neuroinflammatory responses.[316] (**Figure 7**)

Emerging strategies extend beyond amyloid. Reducing intracellular calcium concentration ($[Ca2+]i$) in pericytes has been proposed as a novel target. Early blockade of voltage-gated calcium channels (CaVs) with nimodipine has been reported to improve cerebral blood flow, diminish leukocyte stalling at pericyte loci, and mitigate cerebral hypoxia; CaV inhibition also attenuates Aβ-induced pericyte contraction in human cortical tissue. These findings suggest that lowering pericyte $[Ca2+]i$ in prodromal or early AD may enhance brain energy delivery and support cognitive preservation.[317]

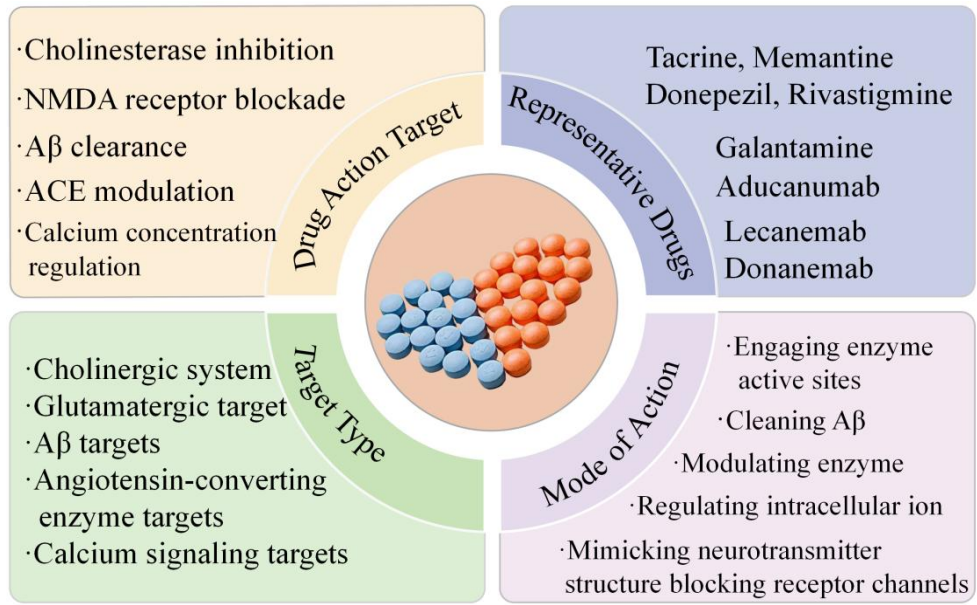

**Figure 7** Alzheimer's Disease: Legacy Medications and Their Role in Therapeutic Evolution

In parallel, immunomodulatory approaches are under investigation. Augmenting angiotensin-converting enzyme (ACE) expression specifically in microglia in 5xFAD mice reduced cerebral Aβ plaque burden, protected vulnerable neurons and excitatory synapses, and rescued learning and memory deficits.[318] This work elucidates an ACE-mediated enhancement of microglial immune function and highlights a potential platform for cell-based or gene-targeted therapies in AD.

Beyond pharmacologic innovation, advances in gene editing, biophysical neuromodulation, and targeted antibody technologies have introduced transformative concepts for Alzheimer's disease (AD) therapy. Gene editing, in particular, has matured rapidly. APOE4—the strongest common genetic risk factor for AD—confers a three- to fivefold increase in risk relative to APOE3 and thus represents a central target.

Two principal strategies are under active investigation:

- Risk reversal via gene replacement. For example, the Cornell AAVrh.10hApoE2 program employs an adeno-associated viral vector to deliver the protective APOE2 allele to the central nervous system.[319] Phase I trials have demonstrated safety, suggesting feasibility for prophylaxis in high-risk individuals.
- In situ gene correction using CRISPR–Cas9. Investigators at MIT have explored editing APOE4 to APOE2-like sequences at key loci, reducing amyloid-β (Aβ) deposition in cellular models. Complementing these efforts, a 2024 Nature study showed that APOE4 activates ACSL1 in microglia, driving lipid droplet accumulation and exacerbating Aβ and tau pathology.[320] This identifies lipid-metabolism targets (e.g., ACSL1) as potential adjuncts for gene-based interventions to improve cognitive outcomes.

Translational hurdles remain substantial, including efficient CNS-targeted delivery, minimization of off-target edits, and rigorous long-term safety evaluation.

Biophysical interventions modulate neural activity and synaptic plasticity using physical energy and offer noninvasive or minimally invasive therapeutic avenues. Deep brain stimulation (DBS) directed at the basal forebrain cholinergic system and the fornix–hippocampal circuit has,[321] in preclinical models, enhanced brain-derived neurotrophic factor (BDNF) release and supported hippocampal neuron survival.[322,323] Transcranial magnetic stimulation (TMS) can adjust excitability within prefrontal–hippocampal networks and has been associated with improvements in working memory in mild AD.[324,325]

The combination of transcranial focused ultrasound with intravenously administered microbubbles has emerged as a promising, non-invasive strategy for treating Alzheimer's disease (AD).[238] The core mechanism involves the targeted application of low-intensity ultrasound waves to specific brain regions, which causes the circulating microbubbles to oscillate.[326] This oscillation mechanically interacts with the endothelial cells of the blood-brain barrier (BBB), transiently and reversibly disrupting its tight junctions. This temporary opening of the BBB facilitates the clearance of neurotoxic amyloid-beta and tau aggregates from the brain into the bloodstream and may also enhance the delivery of therapeutic agents.[327] Preclinical studies have demonstrated significant reduction in pathology and improvement in cognitive function,[327] positioning this approach as a potent dual-function therapy that targets both the pathology removal and drug delivery challenges in AD.

Emerging as a novel neuromodulation approach, 40Hz sensory stimulation—typically delivered via flickering light, auditory tones, or combined audiovisual stimuli—aims to restore impaired gamma oscillations in the brain, which are thought to be crucial for cognitive functions.[328,329] Pioneering preclinical studies demonstrated that driving neural activity at this frequency can reduce pathogenic amyloid-beta and tau protein levels,[96,330] enhance microglial phagocytosis, and improve synaptic function in mouse models of Alzheimer's disease (AD).[331-333] The proposed mechanism involves the entrainment of neuronal networks to synchronize at the gamma frequency,[334] which may promote the brain's innate clearance mechanisms.[335] While initial small-scale clinical trials have reported that this non-invasive intervention is safe and can modulate brain activity and connectivity,[336] its definitive impact on cognitive decline and AD pathology in humans remains an active area of investigation. Ongoing, larger studies are crucial to validate its therapeutic potential and elucidate the underlying molecular and physiological pathways.

These technologies circumvent many pharmacokinetic and systemic side effects of drugs but raise open questions requiring large-scale clinical validation: personalization and optimization of stimulation parameters, durability of therapeutic effects, and mechanisms of synergy with pharmacologic and biologic therapies.

Monoclonal antibodies directed against Aβ and tau constitute the leading paradigm in Alzheimer's disease (AD) immunotherapy. Lecanemab, approved in 2023, demonstrably slows cognitive decline by enhancing Aβ clearance, though its use is constrained by adverse events such as amyloid-related imaging abnormalities (ARIA). Contemporary research is advancing along three fronts:

- Bispecific antibody engineering: constructs simultaneously targeting Aβ oligomers and tau aggregates have achieved dual-pathology reduction in murine models.
- Delivery optimization: nanocarriers and blood–brain barrier–penetrating peptides are being explored to increase central nervous system exposure while limiting peripheral distribution.[337]
- Early-intervention strategies: preventive trials in asymptomatic APOE4 carriers aim to intercept pathogenesis prior to clinical manifestation.[337]

Persistent challenges include high manufacturing costs[338] and heterogeneity in patient response,[339] underscoring the need for bioengineering innovations and biomarker-guided precision medicine.

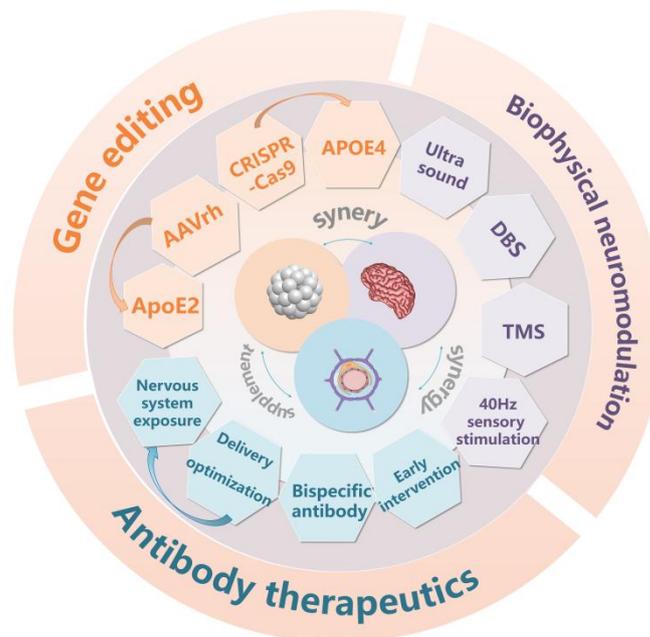

**Figure 8** Integrative Framework of Alzheimer's Disease Interventions: Etiology, Circuitry, Molecular Pathology, and Multimodal Therapeutic Strategies

Collectively, gene editing, biophysical neuromodulation, and antibody therapeutics offer complementary avenues aligned with etiology, circuitry, and molecular pathology, respectively. Gene editing[340] holds promise for durable risk modification via one-time interventions, such as precisely editing pathogenic genes (such as APOE4) using techniques like CRISPR-Cas9, or utilizing AAV vectors to deliver protective genes (such as APOE2);[341] biophysical approaches[342] explore the preservation of fundamental brain functions by targeting core mechanisms of synaptic plasticity, such as employing genetically encoded engineered proteins (GEEP) to enhance synaptic plasticity in living human neurons; antibody therapies aim to achieve precise clearance of pathogenic proteins through meticulously designed molecular strategies.[342] Beyond these, integrated strategies that combine technological strengths are also under development. One team has devised a strategy centered on synaptic repair and neuroprotection, developing an antibody drug that diverges from direct pathogen removal.[343] This approach specifically addresses the synaptic loss and neuronal death characteristic of advanced disease stages, and exhibits a synergistic effect when co-administered with Aβ-clearing therapeutics. As these technologies mature and mechanistic insights deepen, AD management is poised to transition from symptomatic control to genuine disease modification, offering renewed therapeutic prospects for patients.

**6. Conclusion and outlook**

Alzheimer's disease (AD) remains among the most intractable challenges in contemporary medicine. Although substantial progress has been made in elucidating its central pathobiology—principally the amyloid-β (Aβ) and tau cascades—translation into consistently effective therapeutics has proved difficult. Recurrent failures in late-stage Aβ-targeted trials highlight the disorder's complexity, likely reflecting factors such as the high pathological burden at symptomatic stages (potentially beyond reversibility), substantial interindividual heterogeneity, and contributions from additional, non-amyloid/tau mechanisms.

Two paradigm shifts are reshaping AD research and development. First, there is increasing consensus that therapeutic intervention should be initiated earlier—ideally in preclinical or prodromal phases—prior to widespread, irreversible neurodegeneration. Achieving this objective requires scalable, reliable biomarkers, with blood-based assays poised to enable population-level screening and earlier diagnosis. Second, emerging computational technologies, particularly artificial intelligence (AI) and machine learning, are positioned to transform the field. AI-driven platforms can expedite drug discovery by virtually screening large chemical spaces, predicting efficacy and toxicity, and uncovering novel targets within complex biological networks. In parallel, AI methods can integrate multimodal datasets—neuroimaging, genetics, and fluid biomarkers—to define biologically coherent AD subtypes, thereby improving trial stratification and advancing personalized therapeutic strategies.

In conclusion, the therapeutic journey in Alzheimer's disease is evolving from a focus on symptomatic management towards a new paradigm defined by preemptive intervention and combinatorial precision. This paradigm shift is powered by a coherent framework: The widespread adoption of the AT(N) biomarker framework now enables the biological definition of AD years before clinical dementia. This early detection, particularly with the advent of accessible blood-based assays, is not an end in itself but the critical first step. It allows for the mapping of an individual's specific pathological network—their unique constellation of Aβ, tau, neurodegeneration, and neuroinflammatory activity. The complexity of these multi-dimensional data necessitates the application of artificial intelligence, which can deconvolute this heterogeneity to identify biologically coherent subtypes and predict disease trajectories. This AI-driven stratification, in turn, directly informs the selection of personalized, multi-target combination therapies. Thus, a patient with prominent Aβ and tau pathology might receive an anti-Aβ antibody coupled with a tau aggregation inhibitor, while another with a dominant neuroinflammatory profile might be steered toward an immunomodulator like a TREM2 agonist. This dynamic strategy—from early biomarker detection to AI-guided combinatorial precision—finally provides the necessary toolkit to disrupt the self-reinforcing pathological network of AD. It transforms the management of this disease from a passive response to an active, tailored intervention, offering the genuine prospect of altering the course of this challenging disease.